\documentclass[submission,copyright,creativecommons]{eptcs}

\providecommand{\event}{EPTCS journal, work presented at TFPIE} 
\usepackage{underscore}           

\usepackage[T1]{fontenc}
\usepackage[utf8]{inputenc}

\usepackage{graphicx}
\usepackage[T1]{fontenc}
\usepackage{lmodern}
\usepackage{amsmath}
\usepackage{mathpartir}

\usepackage{lstwhy3}
\usepackage{xcolor}
\definecolor{dkgray}{rgb}{.4,.4,.4}
\definecolor{dkred}{rgb}{.4,0,0}
\definecolor{dkgreen}{rgb}{0,.4,0}
\definecolor{dkblue}{rgb}{0,0,.4}

\lstdefinestyle{why3} {
    frame=single,
	language=why3, 
	basicstyle=\ttfamily\scriptsize,
	numbers=left,
	tabsize=4,
	xleftmargin=13pt
}

\lstdefinestyle{why3_inline} {
	language=why3, 
	basicstyle=\ttfamily\footnotesize
}

\usepackage{breakurl}             
\usepackage{underscore}           

\title{Introducing Certified Compilation in Education \\ by a Functional Language Approach}
\author{Per Lindgren
\institute{Luleå University of Technology}
\email{per.lindgren@ltu.se}
\and
Marcus Lindner
\institute{Luleå University of Technology}
\email{marcus.lindner@ltu.se}
\and
Nils Fitinghoff
\institute{Luleå University of Technology}
\email{nilfit-3@student.ltu.se}
}
\def\titlerunning{Introducing Certified Compilation in Education by a Functional Language Approach}
\def\authorrunning{P. Lindgren, M. Lindner \& N. Fitinghoff}

\begin{document}
\maketitle  

\begin{abstract}
Classes on compiler technology are commonly found in Computer Science curricula, covering aspects of parsing, semantic analysis, intermediate transformations and target code generation. This paper reports on introducing certified compilation techniques through a functional language approach in an introductory course on Compiler Construction. Targeting students with little or no experience in formal methods, the proof process is highly automated using the {\tt Why3} framework. Underlying logic, semantic modelling and proofs are introduced along with exercises and assignments leading up to a formally verified compiler for a simplistic imperative language.

This paper covers the motivation, course design, tool selection, and teaching methods, together with evaluations and suggested improvements from the perspectives of both students and teachers.   
\end{abstract} 

\section{Introduction}
Software and software correctness play an undoubtedly increasing role in our society. Correctness of any software application typically relies on the correctness of the compiler at hand, where miscompilation may introduce severe and hard to find errors. 

Over the last decades, formal methods have been gaining momentum in the field, see e.g., the seminal work on CompCert C\cite{Leroy-Compcert-CACM} and LLVM verification \cite{ZhaoNMZ12}. Besides correctness guarantees, the adoption of formal methods forces rigorous semantic modelling and specifications from input language throughout the compilation process. Thus, taking the outset of formal methods into a Compiler Construction course may bring a deeper understanding of the principles of compilation techniques with additional insight into modelling and proofs of programs generally applicable to high assurance application development. So, could the concepts of certified compilation be brought into education, targeting students without prior exposure to formal methods?

A challenging task no doubt, Compiler Construction in its own right covers a vast field, and adopting formal methods without prior experience needs at least a fair introduction. At hand we have in total 9 weeks of half time studies (7.5 ECTS credits\footnote{European Credit Transfer System.}), so topics covered need to be carefully selected. Additionally, exercises and labs should be designed as to maximize learning outcome and motivate students to put in the work needed to gain deeper understanding.

In Section \ref{sec:design} we discuss and motivate the selection of tools and teaching approach, while Section \ref{sec:classes} details lectures and exercises. Section \ref{sec:results} discusses experiences gained from the first installment of the course in 2016. Here we review students' impressions and give a teacher's view. Furthermore, we discuss and detail ongoing improvements to the course (to be given fall of 2018). In Section \ref{sec:related} we review related work, followed by Section \ref{sec:conclusions} where we summarize our contributions.

\section{Course Design}\label{sec:design}
The course in Compiler Construction (D7011E) has been given on a bi-annual basis over the last two decades as an elective course to students of the Computer Science (CS) program at Luleå University of Technology. The course has established a good reputation and typically attracts some 20 students for each installment. For 2016 we decided on reshaping the course, not due to shortcomings of the D7011E course per-se but rather a lack of formal methods in the CS curriculum.

So why not a dedicated formal methods course? Well, the CS curriculum at LTU is already stacked, and introducing a new course would call for removing another. This could have been an option, we have a course on Formal Languages and Theory of Computation D7006E that interleaves with D7011E on a bi-annual basis (which traditionally attracts only a handful of students). However, bi-annual courses pose several problems. Firstly, from a teaching perspective the lack of continuity has clear disadvantages, and secondly, maintaining two interleaved courses poses twice the burden. From a student's perspective, bi-annual courses might be hard to squeeze into their studies (as there are more selectable courses than open slots in the CS program). Thus, replacing D7006E with a dedicated course in formal methods would clearly run the risk of attracting very few students (or cause migration from the D7011E course). 

Hence, we opted to reshape the existing Compiler Construction syllabus from a formal methods outset and incorporate selected topics of D7006E and eventually offer D7011E on a yearly basis after phasing out D7006E from the curriculum. 

\subsection{Course Aims}
Looking to the course aims of prior installments we find the following:  

\fbox{\begin{minipage}{0.9\textwidth}

\paragraph{Course Aim (taken from the official course syllabus prior to 2016)} \hspace{0pt} \\
The student shall be able to
\begin{itemize}
\item Demonstrate the ability to identify and formulate compilation of a high-level programming language into executable machine code as a multi-phase translation process.

\item Demonstrate the ability to implement a compiler for a non-trivial language using appropriate methods.

\item Demonstrate the ability to present and discuss the technological solutions chosen for such an implementation in writing, in an international context.

\item Demonstrate a considerable degree of specialized knowledge in the theoretical foundations of compiler technology.

\item Demonstrate the competence and skill to systematically use proven tools for compiler construction.

\item Demonstrate the ability to analyze and critically evaluate different aspects of modern high-level languages on the basis of their underlying implementation techniques.   
\end{itemize}
\end{minipage}}

Our ambition was to stay with the (already established) course aim and adopt formal methods to improve understanding on the theoretical foundations of compiler technology, while providing additional learning outcomes to high assurance programming (in and beyond the particular scope of Compiler Construction) by introducing semantic specifications, proof techniques etc. 

However, there is no free lunch, and adopting formal methods requires at least a minimum of background information on logic and methods to deductive reasoning beyond the assumed pre-requisite of discrete mathematics (M0009M, given the first year of CS studies). Moreover, in order to adopt formal methods, our input language needs to be sufficiently simple to allow for semantic modelling and proven compilation. Hence realistically, advanced language features, like object orientation, subtyping, type classes/traits, etc. will be out of reach for modelling and proofs in an introductory course.

On the upside, theoretical concepts like Structural Operational Semantics (SOS) and semantically preserving transformations (at heart of {\it any} language and associated compiler) becomes tangible as being concretely integrated in the students' developments. 
With this at hand, we have the basis for discussing extensions of a simplistic language.

In order to cover both compilation techniques and formal methods in an efficient manner, supporting tooling as well as teaching approach were carefully selected.

\subsection{Tools}
Firstly, a functional programming approach was selected in order to facilitate semantic modelling and reasoning on programs. Secondly, we sought a verification platform with a manageable learning curve. Among possible candidates we opted for {\tt OCaml}\footnote{https://ocaml.org/}, backed by the {\tt Why3}\footnote{http://why3.lri.fr/} platform  for logic reasoning and proofs of programs. 

{\tt OCaml} is a modern ML style functional language with an imperative layer and exception handling (facilitating e.g., I/O programming without introducing monads\footnote{For most of the students, this was their first exposure to functional/declarative programming.}). {\tt OCaml} has a rich standard library and a plethora of supporting tools. 

{\tt Why3} is a platform for deductive program verification, where the {\tt Why} language is used for logic (semantic) specifications, and the {\tt WhyML} language for program implementations. {\tt Why3} integrates to a variety of Satisfiability Modulo Theories (SMT) solvers for discharging Verification Conditions (VCs) generated for {\tt WhyML} programs. While the process is highly automated, some degree of interaction may be required for more complex proof cases. To this end, the user can apply transformations on the generated VC(s), such as splitting, simplification, induction, etc. The {\tt Why3} platform allows extraction of {\tt WhyML} programs into {\tt OCaml}, and thus provides a path for certified programming (given the assumption that the {\tt Why3} platform and SMT solver(s) are correctly implemented).

Thirdly, we opted for the {\tt OCaml} {\tt Menhir}\footnote{http://gallium.inria.fr/\textasciitilde fpottier/menhir/} parser generator, motivated by ease of use (well documented, good error messages, etc.). Finally we decided on the {\tt MIPS} 3k RISC architecture as a target for code generation. In this way we could draw on the benefits of the students already being familiar with the architecture as being introduced in a prior course in Microcomputer Engineering (D0013E). Moreover, we could re-use tooling for the compiler backend ({\tt binutils} assembler and linker) as well as the in-house development {\tt SyncSim} for RTL emulation. 

\subsection{Teaching Approach}\label{sec:teaching}

In this course the students installed the necessary tooling on their self hosted laptops. Students brought their computers to the lectures as well as laboratory sessions. The lectures were designed to interleave the introduction of theoretical concepts with hands on demonstrations. With their laptops at hand, the students could replay the examples during class. In this way, theory and practise could be brought together, with the potential benefit of improved attention. 

Laboratory assignments were designed to focus on underlying principles and concepts. In order to maximize efficiency, the students were given boilerplate code/solutions as a baseline for their own development.

Assignments were designed in an open ended fashion, with a minimal requirement for passing each associated topic (e.g., resolving parsing conflicts, construction of a virtual machine with proof of partial correctness to its SOS specification, etc.), with optional requirements towards higher grades. A high degree of freedom for extending/improving on the assignments were given, e.g., adopting more rigorous models and proofs, and/or putting efforts in implementing further code generation optimization techniques. 

To further motivate the students, the ``best compiler'' was crowned at the end of the course.

\section{Lectures, exercises and assignments}\label{sec:classes}

\subsection{Course design and week by week agenda}

For 2016, the week by week agenda was outlined as follows:  
\begin{enumerate}
    \item[w1] Introduction, tools, lexing and parsing.
    \item[w2] Building your compiler frontend.
    \item[w3] Logic and deductive program verification.
    \item[w4] ``imp'' semantics and building a virtual machine (VM).
    \item[w5] Proving correctness of the VM.
    \item[w6] MIPS backend for ``imp''.
    \item[w7] Hoare logic and verification condition generation for ``imp''.
    \item[w8] Optimal and proven register assignment. 
    \item[w9] Examination (individual grading)
\end{enumerate}

As seen, the agenda is quite dense, giving roughly 2 weeks for the frontend (lexing/parsing/AST generation), 3 weeks for program logic (proof techniques and verification of a VM for ``imp''), and 3 weeks for the compiler backend ({\tt MIPS} assembly code generation), encompassing register allocation and rewriting optimizations.

\subsection{Lecture by lecture breakdown}
Lectures were given in a fairly traditional format, each lecture being two times 45 minutes with a short break. Students were encouraged to bring their laptops to follow and replay examples throughout the lectures. In the following we outline the lecture contents as of 2016.

\begin{enumerate}

  
  \item {\tt Introduction}. In the first lecture we cover course texts (Compilers: principles, techniques and tools \cite{Aho:2006:CPT:1177220}) and Real World {\tt OCaml} \cite{citeulike:14362116}), an overview of Compiler Technology (introducing concepts of lexing, parsing into AST representation, desugaring, semantic analysis, linearisation to SSA, high level optimization, RTL level optimization, and ABI conformance). A special focus was given to miscompilation and correctness (GNU C lexing \cite{DBLP:journals/entcs/GlesnerFJ05}, LLVM SSA optimizations \cite{ZhaoNMZ12}), and the CompCert C certified compiler \cite{Leroy-Compcert-CACM}). The use of compilation techniques outside compilers were covered, such as the general use of parsers, transformations and optimizations. A first informal description of the simple imperative ``imp'' language was introduced.

  
  \item {\tt Tools}. The second lecture covers the tooling involved. The programming language {\tt OCaml}, its package manager {\tt opam} and the suggested Eclipse plugin {\tt OcaIDE}; the LR(1) parser generator {\tt menhir} (producing a shift/reduce parser in {\tt OCaml}); and two verification platforms ({\tt Coq} and {\tt Why3}). In an accompanying tutorial session the students installed the tools on their own (or lended) laptops, with the assistance of a TA.

  
  \item {\tt Compiler Frontend}. The third lecture covers lexical analysis, regular expressions and practical details of the {\tt menhir} parser generator.
  Moreover the lecture introduces EBNF grammars with examples to parse Boolean and arithmetic expressions. The concept of algebraic data types in {\tt OCaml} is introduced. Thanks to the {\it extraction} of {\tt WhyML} models to {\tt OCaml}, the generated parser and extracted code can share the same AST definition. The lecture is accompanied with a lab assignment where the students implement new lexing rules (for {\tt strings}). A parser for ``imp'' (syntax) programs into a common AST (shared with {\tt WhyML}) is provided, however it is ambiguous. The tasks for the students are to identify conflicts and come up with a conflict free grammar with well defined precedence and associativity rules. Ambitious students are encouraged to extend the core ``imp'' language with additional constructs (later to be desugared into the core AST).

  
  \item {\tt Logic}. The fourth lecture introduces logic, and deductive reasoning. Firstly, logic inference is demonstrated on propositional logic using the {\tt Why3} framework. Secondly, the theory of First-Order Logic (FOL) is introduced, with concepts such as {\it logical symbols} (quantifiers, connectives, parentheses, punctuations, and optional equality); {\it non-logical symbols} (predicates, functions/constants, and relations); {\it terms} and {\it expressions}; and finally {\it formulas} with the notion of {\it free} and {\it bound} variables with substitution under Leibniz equality, leading up to {\it sentences} with {\it truth values} and {\it interpretations} of FOL. From there we introduce First-order structures, with notions of {\it validity (Tautology), satisfiability} and {\it logical consequence (implication)}.

  The theoretical concepts are backed with a running example of a theory with an interpretation (Peano numbers) in {\tt Why3}. The example provides a natural outset for introducing deductive reasoning (declaring (inductive) predicates, lemmas, theorems and goals), and discussing proof approaches (e.g., proof by induction).

  The lecture is accompanied with a tutorial and a set of assignments, where the students formulate inductive predicates, learn to shape goals and devise various proofs on Peano numbers using the {\tt Why3} framework (where proof obligations are discharged automatically using external SMT solvers).
  
  
  \item {\tt ``imp'' semantics}. In this lecture we give meaning (semantics) to the syntactic structure of an ``imp'' program. The concepts of {\it small-} and {\it big-step} Structural Operational Semantics (SOS) are introduced along with {\it stores} and {\it configurations}. Rules for variable {\it lookup}, arithmetic/Boolean {\it reductions} (expressions), {\it assignments}, {\it sequencing}, {\it conditionals} and {\it loops} are introduced both in small- and big-step form \footnote{The primitive ``imp'' language does not support functions.}. 
  
  We show how the rules for expression evaluation can be encoded in compliance to the SOS, and prove some simple reduction properties. Rules for commands in ``imp'' are captured in the form of an inductive predicate along with a proof of transitivity. 
  
  The lecture is accompanied by a set of assignments, where the students implement the VM, and prove partial correctness to the specification \footnote{Total correctness would be harder as requiring a well founded termination condition ({\it measure}), which is not straightforwardly obtained in the presence of loops.}. The students adopt a logic {\it view} of the store as a map (key/value pairs) \footnote{Variables are of unbound integer type, Booleans are constants (and cannot be assigned/stored for sake of simplicity).}. Expression evaluation is implemented recursively over the algebraic representation, providing a well founded termination condition. However sequences and loops are not structurally decreasing. To this end, a {\it fueling} parameter is added, decreasing for each iteration (rendering a proof of partial correctness), where the VM is guaranteed to either return with an {\it out of fuel} error or a correct (final) configuration \footnote{For any terminating program there exists a fueling parameter rendering a correct final configuration (as the ``imp'' language at hand does not model execution errors).}.
  
  By backing the store view with a list representation, executable {\tt OCaml} code can be extracted from the proven {\tt WhyML} model \footnote{Maps defined in the {\tt Why3} standard library are theories, without backing implementation, hence useful only in proofs and cannot be extracted into executable code.}. 
 
  The ambitious students are encouraged to find the appropriate fueling parameter (measure) for a given program, by tracing the SOS rules (reductions) leading up to termination. (This is a problem related to execution time analysis of the compiled program to the VM language.)

  
  \item {\tt Hoare Logic}. 
  This lecture covers the generation of proof obligations (verification conditions) under Hoare logic, in a similar fashion as implemented by the {\tt Why3} platform. 
  
  We first review the concepts of preconditions and postconditions and their use to form Hoare triples. The axiom schemata for {\it SKIP}, {\it ASSIGNMENT}, {\it COMPOSITION}, {\it CONDITIONAL}, {\it WHILE} and {\it CONSEQUENCE} are introduced and exemplified. The {\it CONSEQUENCE} rule allows for strengthening the precondition and/or weakening the postcondition. Furthermore we cover the computation for weakest liberal \footnote{Referring to partial correctness.} precondition (WP) for loop free code, together with an approximation of weakest precondition for {\it WHILE}. From there we introduce an algorithm for computing Verification Conditions (VCs), but omit formal proof of  correctness. 
  
  In order to make the WP and VC generation tangible, we encode the algorithms by {\it shallow} embeddings in {\tt Why}. This allows us to demonstrate the WP/VC generation for simple programs in ``imp'' and discharge them as {\it Tasks} using the underlying SMT solvers \footnote{WP/VC formulas are represented by internally generated functions, for which we cannot view the concrete representation. However, we can seek their truth values using the {\tt Why3} platform.}

  
  \item {\tt Stack Machine}. In this lecture we revisit the stack machine (originally defined in Double WP \cite{doublewp}).
  We review the proof approach of the compiler backend from ``imp'' to the stack machine language. In particular we focus on the correctness proofs for arithmetic and Boolean expression compilation, while proof details regarding the command layer are only briefly reviewed due to their complexity \cite{clochard:hal-01094488}.

  
  \item {\tt MIPS Assembly}. In a previous course in Microcomputer Engineering, students are exposed to assembly level programming, including stack memory management. In this course they also get initial experience with compiler generated code (using the {\tt GCC} C compiler), the compilation process and tools like {\tt binutils ld}, and the in-house simulator {\tt SyncSim}.

  We review the compilation and linking process and give guidelines for the use of registers. As assignments, students are to implement (unoptimized) code generation along the lines of the Stack Machine covered in a previous lecture \footnote{The {\tt SyncSim} {\tt MIPS} model does not implement native multiplication, thus the students have the option to generate either an error for input programs containing multiplication, or emulate multiplication (for higher marks)}. The compiler should be able to generate assembly output, and at a minimal pass assembling, linking and testing. Ambitious students are encouraged to model the compiler in {\tt WhyML} and use extraction to obtain executable {\tt OCaml} code.
  
  
  \item {\tt Register Allocation}. In this lecture we review register allocation algorithms for tree expressions\cite{register} (including optimal register allocation), and prove them correct using a similar approach to Double WP\cite{clochard:hal-01094488}. 
  The students are encouraged to use the optimal register allocation integrated in their {\tt MIPS} backend \footnote{This can accomplished either by using extraction or by re-implementation directly in the {\tt OCaml} compiler harness.}

  \item {\tt AST level optimization}. Here we introduce a number of optimization techniques applicable at the AST level, namely for expressions {\it constant reduction} (e.g., $a + 1 + 2 \rightarrow a + 3$ , {\it structural equality} (e.g., $a - a \rightarrow 0$), and {\it dominance} (e.g., $true || _ \rightarrow true$), and trivial {\it dead code removal} for conditionals and loops at command level. The students are assigned to model the transformations in {\tt WhyML} and prove semantic preservation to the specification. The extracted code should be integrated in the compiler and optionally enabled through an invocation parameter ({\tt -O n}, {\tt n} being the optimization level). This allows the students to compare generated code efficiency before and after optimization(s).
  
  \item {\tt General Optimization}. In this lecture we give an overview of optimization goals (speed, footprint, dynamic memory usage, power consumption etc.) and discuss obstacles and challenges with respect to correctness and complexity. Optimization at AST, RTL and post processing (link level optimization) are covered. In particular, the concept of general Control Flow Analysis with Local Optimization is discussed (constant folding/common sub-expression elimination), operator strength reduction, copy propagation, non-trivial dead code elimination. Furthermore, global optimizations based on data flow analysis are briefly covered (introducing the concepts of expressions being {\it defined}, {\it available}, and {\it killed}, allowing further sub-expression elimination and re-use. Moreover, code motion (hoisting), register allocation (using graph coloring) and instruction scheduling (peephole optimization and architecture aware exploitation mechanisms for pipelined and VLIW machines) are mentioned. 
  Ambitious students are encouraged to adopt techniques (e.g. hoisting) into their working compiler.
  
  \item {\tt Coq}. In the final lecture, we turn to the interactive proof assistant Coq. Here we revisit the theory of ``imp'' expression evaluation, and model the semantics in terms of an inductive predicate. We give a functional declaration for the evaluation as a {\it fixpoint} definition, and prove it by explicitly applying the induction principle (yielding two sub-goals, one for the base-case, and one for the inductive case.) While the base case can be immediately discharged (solved) by {\it simplification} and {\it reflexivity}, the inductive goal requires interactively applying various {\it tactics}. The example is sufficiently complex to show the trade-off between {\it automation} (as offered by {\tt Why3}) and {\it interaction} as offered by Coq, (the later with the downside of learning curve and effort, but with the upside of a proof process, where the user is in total control provided an {\it informative} view of the proof state (obligations and assumptions) \footnote{Proof state in {\tt Why3} is displayed as a {\it Task}, which may be hard to decipher and control in detail.}. {\tt Why3} allows verification conditions (including proof context) to be exported to Coq. Thus in case automatic solvers do not suffice, an interactive proof process is possible. 
\end{enumerate}

\subsection{Exercises and Laboratory Assignments}

Exercises and assignments have been designed with alignment to the learning goals in mind. Moreover, we have strived to make the assignments both motivating and challenging, with progression throughout the course, (e.g., proof techniques picked up earlier on can be later re-used and refined). Moreover, students may revisit and improve on prior assignments as they mature and gain knowledge during the course.

\subsubsection{Compiler frontend for ``imp''}
In order to facilitate the learning process, a compiler harness ({\tt cimp} written in {\tt OCaml}) was provided, together with a boilerplate lexer and grammar for the {\tt Menhir} parser generator. 

The compiler harness {\tt cimp}, includes stub code for parsing command line arguments, calls the generated parser, performs basic error handling and reporting etc. In this way, the students could directly dig into working with the lexer/parser, to extend lexing rules for strings, to resolve conflicts, define precedence, and come up with their own syntactic extensions.   

\subsubsection{Virtual machine for ``imp''}
The semantic modelling was adopted from the Double WP\cite{clochard:hal-01094488, doublewp} {\tt Why3} development. Double WP provides a model of a certified compiler from the ``imp'' core language to machine code for a minimalistic stack machine. 

Here the students implemented a virtual machine in {\tt WhyML}, and proved its correctness against the semantic specification (inductive predicate). By adopting extractable data types in {\tt WhyML}, {\tt OCaml} code for the proven implementation was obtained and the code could be straightforwardly integrated into the compiler harness.

\subsubsection{Code transformation and {\tt MIPS} backend}
During the course we also adopted techniques from other developments such as the register allocator for tree expressions\cite{register}, and an in house development of a weakest precondition and verification condition generation for the ``imp'' language (based on a shallow embedding in {\tt Why3}).

In this exercise, the students first implemented a simplistic code generator along the lines of the {\it Stack Machine} covered in lectures. Here the register assignment was static, with subexpressions evaluated on the stack.

However, using the stack is highly inefficient, thus in a second part the students implemented register allocation following the lines of the optimal register allocation algorithm presented. Here, ambitious students were encouraged to implement and prove their register allocator in {\tt WhyML}, while less ambitious students could opt for an {\tt OCaml} implementation. As a final exercise, students were to implement and prove a set of simple code transformations at AST level. The ambitious students were encouraged to implement further optimizations (covered in lectures).

\section{Results and Lessons Learned}\label{sec:results}

\subsection{Students' views, Course Evaluation}
Each course installment is followed by a course evaluation, where the students anonymously and voluntarily fill out a questionnaire. For the 2016 installment,  on average each question was answered by more than half of the students (the course was followed by 20 students in total). A summary of the poll is given below with results in italic \footnote{The complete course evaluation amounts to 6 pages (including free text comments mostly in Swedish) can be obtained from the authors on request.}:

\begin{itemize} 

\item How many hours of study have you on average dedicated to this course per week, including both scheduled and non-scheduled
time?  {\it A majority of the students reported 16-25h.}
\item Self Assessment
\begin{itemize}
    \item I am satisfied with my efforts during the course. {\it Average 4.3 of 6.0.}
    \item I have participated in all the teaching and learning activities in the course. {\it Average 5.0 of 6.0.}
    \item I have prepared myself prior to all teaching and learning activities. {\it Average 3.7 of 6.0. }
\end{itemize}
\item Course aims and content

\begin{itemize}
    \item The intended learning outcomes of the course have been clear.  {\it Average 4.0 of 6.0. }
    \item The contents of the course have helped me to achieve the intended learning outcomes of the course. {\it Average 5.0 of 6.0. }
    \item The course planning and the study guide have provided good guidance. {\it Average 4.1 of 6.0. }
\end{itemize}
\item  Quality of teaching

\begin{itemize}
    \item The teacher’s input has supported my learning. {\it Average 5.2 of 6.0. }
    \item The teaching and learning activities of theoretical nature have been rewarding. {\it Average 4.6 of 6.0. }
    \item The practical/creative teaching and learning activities of the course e.g. labs, have been rewarding. {\it Average 4.9 of 6.0. }
    \item The technical support for communication, e.g. learning platform, e-learning resources, has been satisfactory. {\it Average 5.3 of 6.0. }
\end{itemize}

\item Course Materials
\begin{itemize}
    \item The materials assigned for the course, e.g. books, lab instructions, presentation frameworks, has supported my learning. {\it Average 3.7 of 6.0. }

\end{itemize}
\item Examination
\begin{itemize}
    \item The examination was in accordance with the intended learning outcomes of the course. {\it Average 4.0 of 6.0. }
\end{itemize}
\item Overall assessment
\begin{itemize}
    \item The workload of the course is appropriate for the number of credits given.{\it Average 5.1 of 6.0. }
    \item Given the aims of the course the level/difficulty of work required has been appropriate. {\it Average 4.9 of 6.0. }
    \item My overall impression is that this has been a good course. {\it Average 4.9 of 6.0. }
\end{itemize}
\end{itemize}

Among the free text comments, a few highlights (translated to English) is given below:

\begin{itemize}
    \item The course gives a good overview of what a compiler needs to do. Proof of correctness was a pleasant surprise.
    \item Easy to contact the teacher (Telegram) to get help/assistance. Challenging assignments.
    \item Good examples covered in lectures, useful to the lab assignments.
    \item Hands-on and theory mixed to reasonable amount. Large degree of freedom to approach and methods used in labs. Functional languages was a blast.
    \item Interesting subject, and interesting labs.
    \item Interesting setup, fun laboratory assignments, good support from teachers.
    \item Code generation (fun and interesting).
    \item Assignments should be easier to find (now as part of lecture notes).
    \item Change the course name to better reflect course content.
    \item Better distinction between lectures and tutorials (new material was brought up on the tutorials, should be moved to lectures).
    \item Lack of good information on Why3, somewhat fragmented feeling of lectures.
    \item More focus on compiler construction, less boilerplate code (ideally none).
\end{itemize}

\subsection{Teacher's View}

\subsubsection{Overall impression}
As the course was being given for the first time, we were satisfied to see that the efforts spent by the students were in line with the 20 hour per week target (the students take two courses in parallel with a total target of 40 hours workload per week). It is a delicate matter to estimate required efforts, to which end we can conclude to have succeeded. Moreover, the participation at lectures, tutorial sessions and labs remained very high throughout the course (as confirmed by the course evaluation), which leads us to believe that the students found the sessions rewarding. Additionally, the technical support using Telegram worked out very well, both as confirmed by the course evaluation and our experiences teaching the course. 

\subsubsection{Teaching material}
At large, the teaching material (slides, lab instructions etc) sufficed to reach the course aims. However, the lack of easily accessible documentation of the {\tt Why3} framework clearly posed some challenges. The lack of a textbook, presenting compilation techniques from a formal methods perspective was also challenging to teaching the course. 

The lecture slides were designed with the aim to give sufficient background without being too verbose. For the most part, we succeeded, but there is still room for further improvements. However slides alone cannot fully replace a textbook on the subject aiming at the novice to formal methods.

\subsubsection{Learning outcomes and assessment}
As regarding the learning outcomes and meeting course goals, we are pleased with the overall effort spent, dedication and creativity shown by the students. Students were allowed to collaborate in groups of two for the assignments and in building a working compiler. However, most of the students chose to make individual solutions and eventually came up with their own compiler in the end. 

Grading instructions were formulated as follows (snippet from the lecture notes).

\fbox{\begin{minipage}{0.9\textwidth}

\paragraph{Evaluation and grading} \hspace{0pt} \\
Evaluation and grading will be based on your efforts and results obtained. Grades are individual, meaning that even if you worked together on labs, be fair to each other, awarding credit where credit is due. You will bring your developments to the exam date, and demonstrate the implemented features, proofs and obtained results to defend your grade.
\begin{itemize}
	\item Grade 3. All mandatory assignments carried out. (See per-assignment criteria)
	\item Grade 4. Grade 3 + documented efforts into further features and proofs. (See per-assignment criteria.) 
	\item Grade 5. Grade 4 + documented in depth understanding and reflection of concepts covered during the course. 
\end{itemize}
Since first time given in this format, no prior assessment of the efforts required is available, to this end a fair grading is only possible by evaluating Your efforts, contributions and results. 
\end{minipage}}

As seen, the assessment criteria for Grades 3 and 4 were quite concrete (as separately specified per-assignment), while for Grade 5 (highest) assessment was based on documented in depth understanding and reflection and thus left more open ended in order to spawn creativity. 

All students were able to successfully meet the goals for Grade 3, while half of the set of students met the criteria for Grade 4 and a handful of students reached and defended the highest grade (by showing skillful adoption of theoretical concepts covered into their own developments).

\subsection{What is next?}
At the time of writing, the 2018 installment of the course is under preparation. We intend to keep the overall format of the course with the following set of improvements:

\begin{itemize}
    \item Moving all development to {\tt Why3 1.0}. The recent release offers a much improved user experience with the accompanying IDE now offering on the fly editing, interactive proof construction and improved SMT solver integration (with easily accessible counterexample support). 
    \item Further clarification of course goals, expected learning outcomes, examination criteria and compiler competition rules.
    \item Split out assignments/labs from lecture material.
    \item Clear separation between lectures and tutorials (some restructuring of course material).
    \item {\it Strengthening the discussion of formal languages (migrated from D7006E)}. 
    \item {\it Introduction of fixed-width integers, bit-level operations and type checking}. 
\end{itemize}

While, as one student suggested ``{\it the compiler should be built from scratch (not based on boilerplate code)}'', this is not going to happen. The reason is twofold. Firstly, the 9 week time frame does not allow for more code development. Secondly, more coding does not necessarily lead to deeper understanding of the underlying concepts. We strongly believe that boilerplate code is beneficial in the context of this type of course, as allowing (and forcing) the students to focus on the central concepts, not implementation details. 

With the move to {\tt Why3 1.0}, we expect the further streamlining of the laboratory work. Firstly, the new IDE facilitates the proof process (user interaction), and secondly we have been able to simplify the boilerplate code to completely eliminate the need for auxiliary library definitions. Furthermore, the problem of non-termination of ``imp'' program evaluation (and execution of VM code) are now treated by the native {\it diverges} contract in {\tt WhyML}, which allow proofs of partial correctness to be achieve without introducing well founded termination conditions through additional fueling parameters. Other changes include clarifications and slide improvements, which we anticipate to further increase course efficiency. 

Altogether, we feel confident that there is room to strengthen the discussion on formal languages (migrated from D7006E) and introduce fixed-width integers, bit-level operations and type checking. While the arithmetic expressions in ``imp'' operate on unbound integers (as reflected by evaluation and VM execution), target code for the MIPS operate either on signed or unsigned integers (both backed by 32-bit representations). 

In the 2016 installment, students used signed MIPS operations causing exceptions on arithmetic overflows, thus ensuring partial correctness to the specification semantics. However, realistic languages typically provide data types and (unsigned) fixed-width operations reflecting the programmer's intent of wrapping/modulo operational semantics without overflow exceptions. Introducing (unsigned) fixed-width data types and operations thus moves the ``imp'' language a small but important step closer to a realistic language (in particular for high assurance embedded applications). Moreover, from a Compiler Construction point of view, by discriminating between signed and unsigned types and operations, students will encounter type checking as a step in the compilation process. For 2018 students will formulate necessary conditions for well typed programs (under the notions of type casting, conversion, and coersion), implement and (optionally) prove their type checker. Additionally, fixed-width data types enables ambitious students to add bit-vector operations (bit-wise not/and/or/xor etc.) to the ``imp'' language and compiler. 

Obtaining tangible results is highly motivating (while obtaining proofs may not be, unless already interested in theoretical aspects). To that end, a second category for {\it best compiler award} will be added, where students are competing for highest assurance compiler (most rigorously designed and proven).

\section{Related/Similar Work}\label{sec:related}
To the best of our knowledge, the course design is unique in introducing both compilation techniques and deductive program verification to students without prior experience in the respective fields.

The {\tt Why3} homepage \cite{why3} lists a number of courses adopting the {\tt Why3} platform in teaching formal methods and program verification.

\begin{itemize}
    
\item Course Proofs of Programs at the Master Parisien de Recherche en Informatique
\item (in Portuguese) Courses Formal methods and Certified Programming at the Universidade da Beira Interior, Portugal
\item (in French) Course Méthodes formelles et développement de logiciels sûrs at the Master Informatique de l'Université de Rennes
\item (in French) Course Programmation de confiance at the Licence Informatique de l'Université de Rennes
\item (in French) Course sémantique des langages, third year of Supelec Engineering School 
\end{itemize}

In comparison, our focus is on the application of formal methods (not formal methods per-se). Hence, we cover theoretical concepts with less depth than would be possible in a dedicated course. Another difference is that we have a {\it
running example} (our compiler), where we directly {\it apply} and {\it integrate} verification techniques throughout the course.

In the context of traditional courses in compiler technology, in comparison we obviously cover less ground. In particular our ``imp'' language is (implicitly) integer only typed at expression level. Hence, for simplicity type checking is done syntactically (at the stage of lexing/parsing). Advanced type system concepts like object orientation, Traits/typeclasses etc. are omitted from the syllabus in our case. (We aim to close the gap, by including a simple type system/typing rules for 2018.) Moreover, backend code optimization is covered in less depth than would be possible in a dedicated class.   

On the other hand, in comparison to traditional compiler classes, the formalization of the structural operational semantics becomes tangible, as concretely used from (language) specification throughout the compilation process. Adopting formal methods brings the dimension of semantic reasoning, and forces the students to address problems in a rigorous manner.

\section{Conclusions and Future Work}\label{sec:conclusions}

In this paper we report on our experiences introducing certified compilation techniques using a functional language approach in an introductory course on Compiler Construction. We have covered the selection of tools, teaching approach and course syllabus (including a lecture by lecture breakdown and an overview of assignments). Moreover we have discussed results and lessons learned together with an overview of related work in the field.

For the next installment (fall 2018), the course format and syllabus will remain unchanged at large. Notable improvements planned are to strengthen the coverage of formal languages, fixed-with data types and operations, type checking and add a new category of most trustworthy compiler for the student competition (in addition to the best code generation award).

\bibliographystyle{eptcs}
\bibliography{references}

\end{document}